\documentclass[acmsmall,screen]{acmart}
\copyrightyear{2025}
\acmYear{2025}
\acmDOI{}

\renewcommand\footnotetextcopyrightpermission[1]{}

\newif\ifdraft
\draftfalse   
\newif\ifanon
\anonfalse

\newcommand{\wb}[1]{\ifdraft{\color{purple}[#1 -- Will]}\fi}
\newcommand{\web}[1]{\wb{#1}}

\usepackage{alltt}
\usepackage{fancyvrb}
\usepackage{fancyhdr}
\usepackage{subcaption}
\usepackage{cleveref}
\usepackage[euler]{textgreek}

\newcommand{\rc}[1]{{\color{gray}#1}}
\newcommand{\hook}{\rc{$\hookrightarrow$}}

\newcommand{\set}{$\mathcal{S}\mathcal{E}\mathcal{T}$}

\DefineVerbatimEnvironment%
  {VerbatimEsc}{Verbatim}
  {commandchars=\\\{\}}

\begin{document}

\title{Beyond \texttt{Cons}: Purely Relational Data Structures}

\author{Rafaello Sanna}
\email{rsanna@g.harvard.edu}
\affiliation{%
  \institution{Harvard University}
  \country{USA}
}
\author{William E. Byrd}
\email{webyrd@uab.edu}
\affiliation{%
  \institution{University of Alabama at Birmingham}
  \country{USA}
}
\author{Nada Amin}
\email{namin@seas.harvard.edu}
\affiliation{%
  \institution{Harvard University}
  \country{USA}
}

\renewcommand{\shortauthors}{Sanna et al.}

\settopmatter{printacmref=false}
\settopmatter{printfolios=true}
\renewcommand\footnotetextcopyrightpermission[1]{}
\pagestyle{fancy}
\fancyfoot{}
\fancyfoot[R]{miniKanren'25}
\fancypagestyle{firstfancy}{
  \fancyhead{}
  \fancyhead[R]{miniKanren'25}
  \fancyfoot{}
}
\makeatletter
\let\@authorsaddresses\@empty
\makeatother

\ifdraft{
\setcounter{tocdepth}{4}
\setcounter{secnumdepth}{4}
\tableofcontents
\listoffigures
}\fi

\newcommand{\proj}{\{Kanren\}}

\begin{abstract}
We present \proj{} (read: set-Kanren), an extension to miniKanren with constraints for reasoning about sets and association lists. \proj{} includes first-class set objects, a functionally complete family of set-theoretic constraints (including membership, union, and disjointedness), and new constraints for reasoning about association lists with shadowing and scoped lookup. These additions allow programmers to describe collections declaratively and lazily, without relying on structural encodings and eager search over representation spaces. The result is improved expressiveness and operational behavior in programs that manipulate abstract data---particularly interpreters---by supporting set equality based on contents, enabling finite failure. We describe the design and implementation of \proj{} in a constraint-enabled miniKanren system and illustrate its use in representative examples.
\end{abstract}

\keywords{relational programming, constraint solving, data-structures, interpreters, synthesis, miniKanren, Scheme}

\maketitle

\thispagestyle{firstfancy}

\section{Introduction}\label{sec:introduction}

Purely functional programming relies on a rich tradition of persistent data structures---immutable maps, sets, queues, and the like---that make large-scale functional programming practical. These data structures have reshaped how functional programmers reason about performance, representation, and correctness, allowing complex algorithms to be expressed cleanly and efficiently, without requiring mutation.

Relational programming has no such tradition. Systems like miniKanren offer a minimal and elegant core for constraint logic programming, but relational code still relies almost entirely on structural encodings of data---typically lists and trees. Abstract collections such as sets or maps are encoded in terms of these structural primitives. This is limiting not only in expressiveness but also in semantics: structural encodings of sets fail to capture extensional equality and make reasoning about failure and backtracking more difficult.

One promising direction is to encode data-structure operations as constraints. Rather than building data structures structurally and recursing over them, a program instead specifies constraints that describe their shape, contents, and relationships. This makes it possible to support lazy evaluation strategies and more declarative control over search. The theory of constraint-logic programming over sets~\cite{dovierNarrowingTheGap, dovierSetsAndCLP} (written CLP(\set{})) is a notable example: it defines a set of logical constraints for reasoning about extensional sets, supporting membership, union, disjointness, and more. We adapt this work to the miniKanren setting, integrating it with miniKanren’s constraint store and fair search strategy. In doing so, we also extend the approach to cover association lists---another common pattern in relational programs, particularly interpreters.

Concretely, we make the following contributions:

\begin{itemize}
\item We adapt CLP(\set{})'s first-class treatment of relational set operations to miniKanren and extend their work to include association lists
\item We demonstrate how such data-structures improve the failure behavior of relational interpreters and other programs
\item We show how such data-structures can be implemented atop a modern miniKanren implementation
\end{itemize}

The implementation of \proj{} is available at \url{https://github.com/rvs314/faster-clpset-minikanren}.

\section{Background and Motivation}\label{sec:motivation}

\subsection{Motivating Example: Programming Language Semantics}\label{sec:motivation:tapl}

As our first motivating example, consider the following definition~\cite{TAPL} of \textlambda{}-calculus terms~(\cref{fig:tapl-term-def}) and an equivalent rendering into miniKanren~(\cref{fig:tapl-term-mk}).

\begin{figure}[h!]
\begin{subfigure}[t]{0.45\textwidth}
Let $\mathcal{V}$ be a countable set of variable names. The set of terms [in the untyped \textlambda{}-calculus] is the smallest set $\mathcal{T}$ such that:

\begin{enumerate}
\item $\texttt{x} \in \mathcal{T}$ for every $\texttt{x} \in \mathcal{V}$
\item if $\texttt{t}_1 \in \mathcal{T}$ and $\texttt{x} \in \mathcal{V}$, then $\lambda \texttt{x} .\texttt{t}_1 \in \mathcal{T}$
\item if $\texttt{t}_1 \in \mathcal{T}$ and $\texttt{t}_2 \in \mathcal{T}$, then $\texttt{t}_1\ \texttt{t}_2 \in \mathcal{T}$
\end{enumerate}
\caption{An intensional least-fixpoint definition of terms in the untyped \textlambda-calculus, from \citet{TAPL} }~\label{fig:tapl-term-def}
\end{subfigure}\hfill\begin{subfigure}[t]{0.5\textwidth}
\begin{VerbatimEsc}
(defmatche (termo obj)
  [(,x)         (symbolo x)]
  [((\textlambda ,x ,t1)) (symbolo x) (termo t1)]
  [((,t1 ,t2))  (termo t1) (termo t2)])
\end{VerbatimEsc}
    \caption{A miniKanren relation which intensionally describes the set of \textlambda-calculus terms with symbols and lists using the \texttt{matche} syntax of \citet{matche}}\label{fig:tapl-term-mk}
  \end{subfigure}
\caption{An intensional definition of \textlambda-calculus terms, written in both prose and in miniKanren}\label{fig:tapl-term}
\end{figure}

The set of valid \textlambda-terms is very naturally expressed in miniKanren, as it is defined \textit{intensionally}: its members are described in terms of sufficient conditions for membership. However, not all problems can be so phrased without issue. For example, consider the following definition (\cref{fig:tapl-free-def}) and its corresponding implementation (\cref{fig:tapl-free-mk}).

\begin{figure}
\begin{subfigure}[t]{0.45\textwidth}
The set of \textit{free variables} of a term \texttt{t}, written $FV(\texttt{t})$, is defined as follows:

\begin{tabular}{lcl}
$FV(\texttt{x})$ & $=$ & $\{ \texttt{x} \}$ \\
$FV(\lambda \texttt{x} . \texttt{t}_1)$ & $=$ & $FV(\texttt{t}_1)\ \backslash\ \{ \texttt{x} \}$ \\
$FV(\texttt{t}_1\ \texttt{t}_2)$ & $=$ & $FV(\texttt{t}_1)\ \cup FV(\texttt{t}_2)$
\end{tabular}
\caption{An extensional definition of the free variable function~($FV$) over \textlambda-calculus terms, from \citet{TAPL}}~\label{fig:tapl-free-def}
\end{subfigure}\hfill\begin{subfigure}[t]{0.5\textwidth}
\begin{VerbatimEsc}
(defmatche (free-varso obj free)
  [(,x ,f) (symbolo x) (singletono x f)]
  [((\textlambda ,x ,t1) ,f)
   (fresh (f1)
     (free-varso t1 f1)
     (subtracto f1 x f))]
  [((,t1 ,t2) ,f)
   (fresh (f1 f2)
     (free-varso t1 f1)
     (free-varso t2 f2)
     (uniono f1 f2 f))])
\end{VerbatimEsc}
\caption{A miniKanren relation which extensionally describes the set of free variables for a given \textlambda-calculus term}~\label{fig:tapl-free-mk}
\end{subfigure}

\caption{An extensional definition of the free variables in a \textlambda-calculus term, written in both prose and miniKanren}
\end{figure}

In this case, Pierce defines the set of free variables with respect to a given term \textit{extensionally}: the members of the set are listed explicitly. It's clear that intensional sets are, from a semantic perspective, at least as expressive as extensional sets. However, extensional sets allow for reasoning in ways that intensional sets do not; for example, a non-membership constraint is trivial on a finite extensional set, but by Rice's theorem, impossible in general for intensional sets. Following Pierce, we use an extensional definition in \cref{fig:tapl-free-mk}, defining the \texttt{free-varso} relation in terms of the abstract set operations \texttt{singletono}, \texttt{uniono}, and \texttt{subtracto}. What would a concrete implementation of these relations look like?

\begin{figure}
\begin{VerbatimEsc}
(defmatche (ino o l)
  [(,o (,o . ,r))]
  [(,o (,f . ,r)) (ino o r)])

(defmatche (singletono o l) [(,o (,o))])

(defmatche (uniono x y x+y)
  [(() ,y ,y)]
  [((,f . ,r) ,y (,f . ,z)) (uniono r y z)])

(defmatche (subtracto s o s-o)
  [(()        ,o ())]
  [((,o . ,r) ,o ,k) (subtracto r o k)]
  [((,f . ,r) ,o (,f . ,k)) (=/= f o) (subtracto r o k)])

(defmatche (subseto l r)
  [(() ,r)]
  [((,f . ,rst) ,r) (ino f r) (subseto rst r)])

(defrel (set== l r)
  (subseto l r)
  (subseto r l))
\end{VerbatimEsc}
\caption{An example implementation of set operations as lists}\label{fig:list-sets}
\end{figure}

The canonical way to implement sets in a functional language is to approximate them as lists. Ordering and element duplication can be kept hidden through an abstract membership operation and a unification function defined in terms of membership. This technique translates naturally to miniKanren, as in \cref{fig:list-sets}. Such an implementation is sound and complete, but has operational issues. For example, consider the task of unifying such a set against a free variable:

\begin{VerbatimEsc}
(run 100 (q1)
  (fresh (q)
    (free-varso '(\textlambda x y) q)
    (set== q q1)))
\hook{} ((y) (y y) (y y y) ... {\it 97 further results})
\end{VerbatimEsc}

The issue that arises is that miniKanren is searching over the space of possible set {\it representations}, rather than searching over the space of possible sets. This incongruity prevents finite failure and leads to poor performance. What we would like is an implementation of set operations which instead hides the underlying representation from both the programmer and the search algorithm.

\subsection{Motivating Example: Relational Environments}\label{sec:motivation:environments}

As a second motivating example, consider the task of representing environments in a relational interpreter, as in \citet{byrdMiniKanrenLiveAndUntagged}. In that paper, the traditional environment-passing interpreter for a Scheme-like language is translated to miniKanren, including the interpreter's handling of environment lookup and binding, which is based on association lists. The core of the interpreter (\cref{fig:evalo}) is defined in terms of two association list operations\footnote{There is an implicit third operation being used here, which is the binding of association lists. A more abstract implementation of environments might abstract away this operation as well. However, for association lists, the binding operation is well-behaved: it does not cause infinite enumeration, unlike \texttt{not-in-envo} and \texttt{lookupo}.}:  \texttt{not-in-envo} and \texttt{lookupo} (used in the second and third conjuncts of the \texttt{conde} clause, respectively). \texttt{(not-in-envo r e)} ensures that a variable \texttt{r} is not bound in an environment \texttt{e}, and \texttt{(lookupo r e v)} ensures a variable \texttt{r} has a binding with value \texttt{v} in an environment \texttt{e}.

\begin{figure}[h]
\begin{alltt}
(defrel (eval-expro expr env val)
  (conde
    ((fresh (rator rand x body env^ a)
       (== `(,rator ,rand) expr)
       (eval-expro rator env `(closure ,x ,body ,env^))
       (eval-expro rand env a)
       (eval-expro body `((,x . ,a) . ,env^) val)))
    ((fresh (x body)
       (== `(lambda (,x) ,body) expr)
       (symbolo x)
       (== `(closure ,x ,body ,env) val)
       (not-in-envo 'lambda env)))
    ((symbolo expr) (lookupo expr env val))))
\end{alltt}
\caption{The core of the interpreter from \citet{byrdMiniKanrenLiveAndUntagged}}\label{fig:evalo}
\end{figure}

\begin{figure}[h]
\begin{alltt}
(defmatche (not-in-envo r e)
  [(,r ())]
  [(,r ((,k . ,v) . ,t)) (=/= k r) (not-in-envo r t)])

(defmatche (lookupo r e v)
  [(,r ((,r  . ,v)  . ,t) ,v)]
  [(,r ((,r0 . ,v0) . ,t) ,v) (=/= r0 r) (lookupo r t v)])
\end{alltt}
\caption{The association list operations implemented in terms of primitive recursion and search}\label{fig:lookupo-lists}
\end{figure}

The canonical implementation of these relations is to define them based on the recursive structure of an association list (see \cref{fig:lookupo-lists}). These implementations are sound and complete, but cause issues similar to those found in \cref{sec:motivation:tapl}; namely, queries against non-ground lists quickly lead to infinitely many results. To illustrate this, consider the following attempt to evaluate a combinator against a non-ground environment:

\begin{VerbatimEsc}
(run 100 (env val)
  (eval-expro `(lambda (x) x) env val))
\hook{} ((() (closure x x ()))
    ((((_.0 . _.1)) (closure x x ((_.0 . _.1))))
     (=/= ((_.0 lambda))))
    ((((_.0 . _.1) (_.2 . _.3)) (closure x x ((_.0 . _.1) (_.2 . _.3))))
     (=/= ((_.0 lambda)) ((_.2 lambda))))
    ... {\it 97 further results})
\end{VerbatimEsc}

As can be seen, we are searching over the space of all {\it representations} of environments which do not bind the name \texttt{lambda}. This enumeration is caused by the \texttt{not-in-envo} constraint applied to the environment in the third clause of the disjunction in \cref{fig:evalo}: in order to ensure that no environment binds \texttt{lambda} as a variable name, we begin enumerating all environments in which the name is not bound\footnote{One folklore solution to this problem is the use of a ``split-environment representation'', in which the keys of the association list are stored redundantly in a separate list. To implement \texttt{not-in-envo}, the free name is prohibited from occurring in the list of keys using \texttt{absento}. This representation `fakes' the behavior of a constraint-based data-structure in terms of an existing constraint. This representation comes with its own host of issues, mainly due to the incongruity between the information stored in the \texttt{absento} constraints and the information stored structurally in the association list.}.

\section{\proj{} Extensions}\label{sec:constraints}

To remain compatible with existing miniKanren programs, all traditional miniKanren types and constraints are valid \proj{} types and constraints. In addition to this base, \proj{} introduces three new extensions: set objects, constraints over sets, and constraints over association lists.

\subsection{Set Objects}\label{sec:constraints:set-objects}

\begin{table}[h]
\begin{tabular}{c|c|c}
\textbf{Name} & \textbf{Scheme Representation} & \textbf{Denotation} \\
\hline
Empty Set     & \verb|#(set)|             & $\emptyset$ \\
\hline
Proper Set    & \verb|#(set (a b ...))|   & $\{ \texttt{a}, \texttt{b}, ... \}$ \\
\hline
Improper Set  & \verb|#(set (a b ...) r)| & $\{ \texttt{a}, \texttt{b}, ... \} \cup \texttt{r}$ \\
\end{tabular}
\caption{Set objects in \proj{}}\label{tab:set-forms}
\end{table}

\proj{} represents sets as tagged Scheme vectors\footnote{Traditionally, miniKanren implementations represent logic variables as Scheme vectors. To distinguish between set objects and logic variables, \proj{} variables are tagged vectors which start with the symbol \texttt{'var}.}. A set object takes one of the three forms listed in \cref{tab:set-forms}:
\textbf{Empty sets}, which contain \textit{no} elements;
\textbf{Proper sets}, which contain \textit{exactly} a known set of elements (up to order and duplication);
and \textbf{Improper sets}, which contain \textit{at least} a known set of elements (up to element order and duplication).
Notice that each set form subsumes the forms before it: set objects may be written in a number of semantically equivalent but syntactically distinct forms. However, when set objects are returned as query results, they are normalized to a canonical form. This form is the smallest member of the set's current equivalence class: that is, the set's elements are conservatively deduplicated and grouped into a single vector when possible, and unnecessary nesting is removed.

Set objects are recognized by the type-constraint \texttt{seto}. Like other type constraints, \texttt{seto} is lazy: it does not force the variable to which it is applied to choose a particular ground value, ensuring only that any chosen value is a valid set object. Additionally, this constraint is recursive: it checks not only that the term to which it is applied is a set, but also that all improper sets' third element (the ``tail'' of the improper set) is itself a set.

Set objects are compatible with existing miniKanren constraints, including unification constraints (\texttt{==}), disunification constraints (\texttt{=/=}), and absence constraints (\texttt{absento}). These operations interpret sets extensionally: two set objects unify if and only if they represent the same set, regardless of representation or ordering. As an example, consider the task of finding all values of \texttt{p} such that the unification $\{ 1, 2, 3 \} \equiv \{ 2, 3 \} \cup \texttt{p}$ holds:

\begin{alltt}
(run* (p) (== `#(set (1 2 3)) `#(set (2 3) ,p)))
\hook{} (#(set (1)) #(set (1 2)) #(set (1 3)) #(set (1 2 3)))
\end{alltt}

\noindent
The resulting list of answers\footnote{Note: The traditional miniKanren invariant that unification succeeds at most once does not hold in \proj{} programs which use sets. \proj{} does, however, guarantee that unification will succeed at most a finite number of times. This is further expanded upon in \cref{subsec:unification}.} includes all the values of $p$ such that $p \in \mathcal{P}(\{1, 2, 3\})$ and $1 \in p$.

\subsection{Set Constraints}\label{sec:constraints:set-constraints}

\begin{table}[h]
\centering
\begin{tabular}{c|c|c}
\textbf{Constraint Signature} & \textbf{Denotation} & \textbf{Definition} \\
\hline
\verb|(!ino el st)|      & $\texttt{el} \notin \texttt{st}$ & Primitive \\
\hline
\verb|(disjo a b)|       & $\texttt{a} \cap \texttt{b} \equiv \emptyset$
                         & Primitive \\
\hline
\verb|(uniono a b c)|    & $\texttt{a} \cup \texttt{b} \equiv \texttt{c}$
                         & Primitive \\
\hline
\verb|(ino el st)|       & $\texttt{el} \in \texttt{st}$
                         & $\exists p.\, \{ \texttt{el} \} \cup p \equiv \texttt{st}$ \\
\hline
\verb|(union+o l r c)|& $\texttt{l} \uplus \texttt{r} \equiv \texttt{c}$ & $\texttt{l} \cup \texttt{r} \equiv \texttt{c} \land \texttt{l} \cap \texttt{r} \equiv \emptyset$ \\ \hline
\verb|(!uniono l r c)|   & $\texttt{l} \cup \texttt{r} \not\equiv \texttt{c}$
                         & \begin{tabular}{ll}
         $\exists n.\,   (n \in \texttt{c} \land n \notin \texttt{l} \land n \notin \texttt{r})\ \lor$\\
\phantom{$\exists n.\,$}$(n \notin \texttt{c} \land (n \in \texttt{l} \lor n \in \texttt{r}))$ \\
\end{tabular}\\
\hline
\verb|(!disjo l r)|      & $\texttt{l} \cap \texttt{r} \not\equiv \emptyset$ & $\exists n.\, n \in \texttt{l} \land n \in \texttt{r}$\\
\hline
\verb|(subseteqo b p)|   & $\texttt{b} \subseteq \texttt{p}$ & $\texttt{b} \cup \texttt{p} \equiv \texttt{p}$ \\
\hline
\verb|(subseto b p)|     & $\texttt{b} \subset \texttt{p}$ & $\texttt{b} \cup \texttt{p} \equiv \texttt{p} \land (\exists n.\, n \in \texttt{p} \land n \notin \texttt{b})$ \\
\hline
\verb|(subtracto l o w)| & $\texttt{l} - \{ \texttt{o} \} \equiv \texttt{w}$ & $\texttt{o} \notin \texttt{w} \land (\texttt{l} \equiv \{ \texttt{o} \} \cup \texttt{w} \lor \texttt{l} \equiv \texttt{w})$
\end{tabular}
\caption{Set constraints in \proj{}} \label{tab:set-constraints}
\end{table}

While some constraints may be written purely in terms of unification, disunification and absence constraints, there are several which cannot be so expressed. As shown by \citet{dovierNarrowingTheGap}, an implementation must provide a functionally complete set of lazy constraints to ensure finite failure. There are many functionally complete operator sets, but \proj{} follows \citet{dovierSetsAndCLP} in defining three primitive operations: non-membership (\texttt{!ino}), disjointedness (\texttt{disjo}) and set-union (\texttt{uniono}). With these, we can define the typical set operators: subsets (\texttt{subseto}), disjoint unions (\texttt{union+o}), and others. \cref{tab:set-constraints} lists each provided constraint and its definition in terms of primitive constraints. For an example of this, consider the following query, which demonstrates how the disjoint union constraint is reified as a disjointedness constraint and a union constraint:

\begin{alltt}
(run* (l r c) (union+o l r c))
\hook (((_.0 _.1 _.2)
     (set _.0 _.1 _.2)
     (\ensuremath{\parallel} (_.0 _.1))
     (\(\cup{}\sb{3}\) (_.0 _.1 _.2))))
\end{alltt}

\subsection{Association List Constraints}\label{sec:constraints:alist-constraints}

\begin{table}[h]
\begin{tabular}{c|c}
\textbf{Name} & \textbf{Denotation} \\
\hline
\verb|(listo n)|  & $\texttt{n} \equiv \verb|'()| \lor (\exists\ a, d.\,\texttt{n} \equiv (\texttt{cons}\ a\ d) \land (\texttt{listo}\ d))$ \\
\hline
\verb|(freeo k l)| & $\texttt{l} \equiv \verb|'()| \lor (\exists\ k_0, v, r.\,\texttt{l} \equiv (\texttt{cons}\ (\texttt{cons}\ k_0\ v)\ r) \land k_0 \neq k \land (\texttt{freeo}\ \texttt{k}\ r))$ \\
\hline
\verb|(lookupo k l v)| & \begin{tabular}{ll}
         $\exists\ k_0, v_0, r.\, \texttt{l} \equiv (\texttt{cons}\ (\texttt{cons}\ k_0\ v_0)\ r) \ \land$\\
\phantom{$\exists\ k_0, v_0, r.\,$}$([k_0 \equiv \texttt{k} \land v_0 \equiv \texttt{v}] \lor [k_0 \neq \texttt{k} \land (\texttt{lookupo}\ \texttt{k}\ r\ \texttt{v})])$ \\
\end{tabular} \\
\end{tabular}
\caption{Association list constraints in \proj{}}\label{tab:alist-constraints}
\end{table}

Rather than operating on a new datatype, the association list constraints of \proj{} act on typical Scheme lists made of \texttt{cons} cells. The reasons for this are twofold: firstly, it allows for interoperation with existing miniKanren code which uses association lists. Secondly, association lists have a natural notion of order, where elements added more recently shadow those that have been added earlier.

To support reasoning about association lists, \proj{} adds three new constraints (as listed in \cref{tab:alist-constraints}): \verb|(listo l)|, which asserts that \texttt{l} is a proper list (it is either null or a pair whose \texttt{cdr} is a proper list); \verb|(freeo k l)|, which asserts that \texttt{l} is a proper association list (a proper list whose elements are all pairs) and that the key \texttt{k} never occurs in the association list; and \verb|(lookupo k l v)|, which asserts that \texttt{l} is a proper list\footnote{Note: the constraint does not force the list to be a proper association list, as \texttt{freeo} does. \texttt{lookupo} only forces the elements \textit{up to the key-value pair to which it corresponds} to be pairs. If no such association exists, it forces all elements to be pairs.} which associates \texttt{k} with \texttt{v}. Because we implement association lists as traditional Scheme lists, we do not require an explicit association extension constraint. The association list \texttt{l} with \texttt{k} mapped to \texttt{v} is written \verb|(cons (cons v k) l)|, as in traditional association lists.

\section{Case Studies}\label{sec:casestudies}

In addition to the motivating examples presented in sections \ref{sec:motivation:tapl} and \ref{sec:motivation:environments}, in this section we present additional simple case studies showing the utility of these constraints.

\subsection{Case Study: Programming Language Semantics}\label{sec:casestudy:tapl}

As discussed in \cref{sec:motivation:tapl}, extensional set operations arise naturally and frequently in programming language semantics, where many relations compute or quantify over sets of variables, types, or evaluation contexts. The \proj{} set API is expressive enough to capture these relations directly. Consider the definition of \texttt{free-varso}, given previously in \cref{fig:tapl-free-mk}. As \proj{} implements all of the abstract operations required by the definition, we can run the prior example without any changes:

\begin{alltt}
(run* (q1)
  (fresh (q)
    (free-varso '(\textlambda x y) q)
    (== q q1)))
\hook{} (#(set (y)))
\end{alltt}

In this case, \proj{} is able to resolve the result of the query to a single set, as sets can be unified against, just like any other object, rather than abstract structures which require a custom unification relation.

\subsection{Case Study: Relational Environments}\label{sec:casestudy:environments}

As discussed in Section~\ref{sec:motivation:environments}, relational interpreters often model environments using association lists, where bindings are introduced at the head of the list and earlier bindings are shadowed by later ones. This representation, when implemented eagerly, creates difficulties for search: constraints such as \texttt{not-in-envo} may trigger enumeration of the entire space of possible bindings.

It should be noted that while both set constraints and association list constraints serve a similar purpose, the behaviors of one family of constraints cannot be expressed using the constraints of the other: for example, the \texttt{not-in-envo} relation must disallow bindings with a particular key, but only when that key appears at the head of the list. Even if we could encode this behavior extensionally, we would still need to reason about shadowing, a fundamentally structural property. In contrast to sets, association lists are ordered collections, and the semantics of lookup and binding depend critically on that ordering.

These constraints \texttt{freeo}\footnote{We chose to name this constraint \texttt{freeo} rather than \texttt{not-in-envo}, as the former implies a wider range of use-cases.} and \texttt{lookupo} are near drop-in replacements for their structural counterparts, but exhibit better operational behavior in the presence of logic variables. As an example, consider the relational interpreter listed earlier (\cref{fig:evalo}). When \texttt{not-in-envo} is implemented structurally, querying for the closure of a variable in a non-ground environment leads to infinite enumeration. In contrast, when \texttt{not-in-envo} is replaced with \texttt{freeo}, the query gives a single answer:

\begin{alltt}
(load "./simple-interp.scm") ; Interpreter from \citet{byrdMiniKanrenLiveAndUntagged}, see \cref{fig:evalo}

(set! not-in-envo freeo)

(run* (env val)
  (eval-expro `(lambda (x) x) env val))
\hook{} (((_.0 (closure x x _.0)) (lst _.0) (free (lambda _.0))))
\end{alltt}

\subsection{Case Study: Remembering Vertices Encountered in a Graph}\label{sec:tablingforpaths}

Another simple use of set constraints is for remembering vertices that have been encountered when computing reachability in a directed graph.
For example, consider these relations taken from Section 12.4 of \citet{byrdthesis}:

\begin{alltt}
(define (patho x y)
  (conde
    ((arco x y))
    ((fresh (z)
       (arco x z)
       (patho z y)))))

(define (arco x y)
  (conde
    ((== 'a x) (== 'b y))
    ((== 'b x) (== 'a y))
    ((== 'b x) (== 'd y))))
\end{alltt}

\noindent
The \texttt{arco} relation specifies edges in a directed graph with vertices \texttt{a}, \texttt{b}, and \texttt{d}: this graph has directed edges from \texttt{a} to \texttt{b}, from \texttt{b} to \texttt{a}, and from \texttt{b} to \texttt{d}.
The \texttt{patho} relation declares that a path exists between vertices \texttt{x} and \texttt{y} if there exists a direct edge between \texttt{x} and \texttt{y} (defined in \texttt{arco}), or if there exists a direct edge from \texttt{x} to some vertex \texttt{z}, and there exists a path from \texttt{z} to \texttt{y}.

Since the graph defined in \texttt{arco} contains a cycle, there are infinitely many paths starting from vertex \texttt{a}:

\begin{alltt}
(run 10 (q) (patho 'a q))
\hook{} (b a d b a d b a d b)
\end{alltt}

If we want to avoid duplicate paths produced by the cycle, we can use \texttt{path-tabledo}, which keeps track of the set of vertices that have already been encountered, and only produces direct or indirect paths that reach new vertices.

\begin{samepage}
\begin{alltt}
(define (path-tabledo x y table)
  (conde
    ((!ino y table)
     (arco x y))
    ((fresh (z)
       (arco x z)
       (!ino z table)
       (fresh (table^)
         (== `#(set (,z) ,table) table^)
         (path-tabledo z y table^))))))
\end{alltt}
\end{samepage}

\noindent
Using \texttt{path-tabledo}, starting with an empty table of seen vertices, only a finite set of paths is reachable from \texttt{a}:

\begin{alltt}
(run* (q) (path-tabledo 'a q '#(set)))
\hook{} (b a d)
\end{alltt}

We can easily generalize our path finding relations by replacing the hard-coded \texttt{arco} relation with a set containing directed edges of the form \texttt{(x -> y)}.

\begin{alltt}
(define (path-with-edgeso x y edge-set)
  (conde
    ((ino `(,x -> ,y) edge-set))
    ((fresh (z)
       (ino `(,x -> ,z) edge-set)
       (path-with-edgeso z y edge-set)))))

(run 10 (q) (path-with-edgeso 'a q '#(set ((a -> b)
                                           (b -> a)
                                           (b -> d)))))
\hook{} (b a d b a d b a d b)
\end{alltt}

And, once again, we can keep a set of vertices that have already been encountered to finitize the paths starting from \texttt{a}.

\begin{alltt}
(define (path-with-edges-tabledo x y edge-set table)
  (conde
    ((!ino y table)
     (ino `(,x -> ,y) edge-set))
    ((fresh (z)
       (ino `(,x -> ,z) edge-set)
       (!ino z table)
       (fresh (table^)
         (== `#(set (,z) ,table) table^)
         (path-with-edges-tabledo z y edge-set table^))))))

(run* (q)
    (path-with-edges-tabledo
        'a q '#(set ((a -> b) (b -> a) (b -> d))) '#(set)))
\hook{} (b a d)
\end{alltt}

\noindent
Since we are now representing the graph as a set of edges, we can specify starting and ending vertices and synthesize directed graphs that connect those vertices:

\begin{alltt}
(run 3 (q) (path-with-edges-tabledo 'a 'b q '#(set)))
\hook{} 
((#(set ((a -> b)) _.0)
  (set _.0))
 (#(set ((_.0 -> b) (a -> _.0)) _.1)
  (=/= ((_.0 b)))
  (set _.1))
 (#(set ((_.0 -> _.1) (_.1 -> b) (a -> _.0)) _.2)
  (=/= ((_.0 _.1)) ((_.0 b)) ((_.1 b)))
  (set _.2)))
\end{alltt}

\subsection{Case Study: metaKanren with a Set-based \texttt{run} Interface}\label{sec:metaKanrenrun}

There have been several implementations of miniKanren or microKanren inside of miniKanren.
One such implementation is \emph{metaKanren} \cite{metaKanren}, which provides an \texttt{eval-programo} relation capable of running miniKanren code inside of miniKanren:

\begin{alltt}
(run* (q)
  (eval-programo
   `(run* (z)
      (disj (== z 1)
            (== z 2)))
   q))
\hook{} ((1 2))
\end{alltt}

Because \texttt{eval-programo} is a relation, the user can specify the value of the \texttt{run} or \texttt{run*} being evaluated by \texttt{eval-programo}, and place logic variables in the miniKanren expression being evaluated by \texttt{eval-programo}, thereby synthesizing miniKanren code:

\begin{alltt}
(run 1 (q)
  (eval-programo
   `(run* (z)
      (disj (== ,q 1)
            (== z 2)))
   '(1 2)))
\hook{} (z)
\end{alltt}

\noindent
Unfortunately, this technique requires correctly specifying as the second argument to \texttt{eval-programo} not just the results produced by the first argument to \texttt{eval-programo}, but also the \emph{order} in which those results are produced.
For example, if we were to swap the 1 and 2 in the list (1 2) above, the resulting query

\begin{alltt}
(run 1 (q)
  (eval-programo
   `(run* (z)
      (disj (== ,q 1)
            (== z 2)))
   '(2 1)))
\end{alltt}

\noindent
would diverge.
This sensitivity to the order of the results makes it difficult to express interesting synthesis queries using \texttt{eval-programo}.
However, we can avoid this problem by modifying metaKanren so that \texttt{eval-programo} takes a \emph{set} of results as its second argument, instead of an ordered list.

To make this change, we need to update three definitions in the metaKanren implementation: \texttt{take-allo}, \texttt{take-no}, and \texttt{reifyo}.
These changes are shown in figure~\ref{fig:metaKanren}, which shows the list-based and set-based definitions of \texttt{take-allo}, \texttt{take-no}, and \texttt{reifyo} side-by-side.

\begin{figure}[h!]
\begin{subfigure}[t]{0.4\textwidth}
\begin{VerbatimEsc}
(define (take-allo $ s/c*)
  (fresh ($1)
    (pullo $ $1)
    (conde
      ((== '() $1) (== '() s/c*))
      ((fresh (a res $d)
         (== `(,a . ,$d) $1)
         (== `(,a . ,res) s/c*)
         (take-allo $d res))))))

(define (take-no n $ s/c*)
  (conde
    ((== '() n) (== '() s/c*))
    ((=/= '() n)
     (fresh ($1)
       (pullo $ $1)
       (conde
         ((== '() $1) (== '() s/c*))
         ((fresh (n-1 res a d)
            (== `(,a . ,d) $1)
            (== `(,n-1) n)
            (== `(,a . ,res) s/c*)
            (take-no n-1 d res))))))))

(define (reifyo s/c* out)
  (conde
    ((== '() s/c*) (== '() out))
    ((fresh (a d va vd)
       (== `(,a . ,d) s/c*)
       (== `(,va . ,vd) out)
       (reify-state/1st-varo a va)
       (reifyo d vd)))))
\end{VerbatimEsc}
\caption{Original list-based definitions of {\tt take-allo}, {\tt take-no}, and {\tt reifyo} from metaKanren.}~\label{fig:metaKanrena}
\end{subfigure}\hfill\begin{subfigure}[t]{0.55\textwidth}
\begin{VerbatimEsc}
(define (take-allo $ s/c-set)
  (fresh ($1)
    (pullo $ $1)
    (conde
      ((== '() $1) (== '#(set) s/c-set))
      ((fresh (a res $d)
         (== `(,a . ,$d) $1)
         (== #`(set (,a) ,res) s/c-set)
         (take-allo $d res))))))

(define (take-no n $ s/c-set)
  (conde
    ((== '() n) (== '#(set) s/c-set))
    ((=/= '() n)
     (fresh ($1)
       (pullo $ $1)
       (conde
         ((== '() $1) (== '#(set) s/c-set))
         ((fresh (n-1 res a d)
            (== `(,a . ,d) $1)
            (== `(,n-1) n)
            (== #`(set (,a) ,res) s/c-set)
            (take-no n-1 d res))))))))

(define (reifyo s/c-set out)
  (conde
    ((== '#(set) s/c-set) (== '#(set) out))
    ((fresh (a d va vd)
       (== `#(set (,a) ,d) s/c-set)
       (== `#(set (,va) ,vd) out)
       (!ino va vd)
       (reify-state/1st-varo a va)
       (reifyo d vd)))))
\end{VerbatimEsc}
\caption{Updated definitions of {\tt take-allo}, {\tt take-no}, and {\tt reifyo} that produce sets of values (or sets of reified values).}\label{fig:metaKanrenb}
\end{subfigure}
\caption{Modifications to metaKanren~\cite{metaKanren} allowing \texttt{eval-programo} to take a \emph{set} of results as its second argument, instead of an ordered list.  To make this change we need to update three definitions in the metaKanren implementation: \texttt{take-allo}, \texttt{take-no}, and \texttt{reifyo}.  The original definitions of \texttt{take-allo} and \texttt{take-no} each produce a list of values \texttt{s/c*} from a stream \texttt{\$}, while the original list-based \texttt{reifyo} definition produces a reified list of values \texttt{out} from a non-reified list of value \texttt{s/c*} (\ref{fig:metaKanrena}). Changing these three definitions to produce sets instead of lists is straightforward (\ref{fig:metaKanrenb}).  The \texttt{(!ino va vd)} constraint in the set-based \texttt{reifyo} is needed to prevent the reifier from producing an unbounded number of duplicate answers. \web{why is this???}}\label{fig:metaKanren}
\end{figure}

We can now modify one of the queries from \citet{metaKanren} to take a set:

\begin{alltt}
(run-unique*\footnote{The \texttt{run-unique*} operator is functionally similar to the \texttt{run*} of traditional miniKanren, with the exception that it removes results which are syntactically identical. For why this is necessary, see \cref{sec:limitations:duplicates}.} (count)
  (eval-programo
   `(run ,count (z)
      (disj (== z 1)
            (== z 2)))
   '#(set (1 2))))
\hook{} 
(((()))
 (((_.0)) (=/= ((_.0 ())))))
\end{alltt}

\noindent
which produces the same result as:

\begin{alltt}
(run-unique* (count)
  (eval-programo
   `(run ,count (z)
      (disj (== z 1)
            (== z 2)))
   '#(set (2 1))))
\end{alltt}

\section{Implementation}\label{sec:implementation}

An initial version of \proj{} was implemented from scratch~\cite{clpset-miniKanren}, following the constraint rewriting procedures of~\citet{dovierSetsAndCLP}. The current version of \proj{} is implemented as a fork of the faster-miniKanren project~\cite{faster-miniKanren}. The faster-miniKanren project implements a number of optimizations, including:

\begin{itemize}
\item An optimized constraint and lookup store
\item A safe mutation shortcut for non-backtracked variables
\item Attributed variables~\cite{AttributedVariables} for efficient constraint propagation
\item A lazy disunification implementation similar to a two-watched literal scheme~\cite{Chaff, GRASP}
\end{itemize}

In many ways, faster-miniKanren has become the standard implementation of miniKanren for non-trivial use-cases due to its comprehensive collection of constraints and efficient implementation. However, faster-miniKanren makes a number of assumptions about miniKanren that are violated by \proj{}. We relax these assumptions, allowing for more diverse constraints and solvers. Some of the major changes made by \proj{} are:

\begin{itemize}
\item The generalization of unification to arbitrary goals (\cref{subsec:unification})
\item The use of a first-order search tree to enable efficient disunification (\cref{subsec:disunification})
\item The ``virtualization'' of the \texttt{absento} constraint (\cref{subsec:virtualizing-absento})
\end{itemize}

We discuss these changes below.

\subsection{Unification with Structural Constraints}\label{subsec:unification}

Traditional miniKanren unification (as presented in \citet{byrdthesis} and \citet{reasoned2}) is implemented in terms of a \texttt{(unify u v s)} procedure, which, for some substitution \texttt{s}, returns \texttt{\#f} if \texttt{u} and \texttt{v} cannot be unified, or a new substitution $\hat{\texttt s}$, an extension of \texttt{s} which binds \texttt{u} and \texttt{v} to their most general unifier (MGU)\footnote{For performance reasons, unification in faster-miniKanren and \proj{} also returns information in addition to the MGU---namely, the set of variables which have been bound---in order to perform efficient constraint propagation.}. This structure makes two assumptions: first, that there exists a unique MGU; and second, that unification does not require the usage of non-unification constraints (and can be implemented in terms only of the substitution). These are reasonable assumptions for the dialect of miniKanren implemented by faster-miniKanren and for many extensions of that language, but are not true in general: \proj{}, for example, breaks both of them.

Firstly, there is not necessarily a unique MGU for set unification in the theory of CLP(\set{}). We saw an instance of this in \cref{sec:constraints:set-objects}, when we attempted the unification $\{ 1, 2, 3 \} \equiv \{2, 3 \} \cup \texttt{p}$. In that case, there existed ground and mutually exclusive bindings for the query variable \texttt{p} that satisfy the unification constraint. \proj{} supports such bindings by generalizing the return type of \texttt{unify} to a stream of unification results. Although it cannot guarantee a single result, \proj{} ensures that there will be at most a finite number of results, as \proj{} only reasons about finite sets.

Secondly, CLP(\set{}) requires that unification be able to perform non-unification constraints such as type assertions and existential generalization. For example, consider the following query:

\begin{alltt}
(run* (p q)
  (== `#(set (1) ,p) `#(set (2) ,q)))
\hook{} (((#(set (2) _.0) #(set (1) _.0)) (set _.0)))
\end{alltt}

\noindent
Despite only performing a unification, the fresh variable \texttt{\_.0} was introduced and is assigned the \texttt{set} type constraint. In order to include arbitrary constraints as part of unification, \proj{} generalizes the \texttt{unify} procedure into one which takes a state object and returns a stream of potential future states, as with other miniKanren goals. While a conceptually simple change, these changes in unification affect the performance and correctness of many other parts of the implementation.

\subsection{Recovering Efficient Disunification with a First-Order Representation}\label{subsec:disunification}

The optimized disunification implementation used by faster-miniKanren relies on the assumption that unification of two terms implies a conjunction of unifications to variables. By DeMorgan's law, disunification of those terms then corresponds to a disjunction of disunifications to variables. Rather than forking the search for each disjunct, faster-miniKanren stores the disjunction explicitly. This reification enables an efficient propagation strategy similar to the two-watched literal technique used in SAT solvers~\cite{Chaff, GRASP}: only one variable per disjunct needs to be monitored, and the constraint is reevaluated only when \textit{that} variable changes.

In the presence of CLP(\set{}) constraints, this decomposition no longer holds. As seen in \cref{subsec:unification}, set unification may introduce fresh variables and constraints beyond those expressible as substitutions. Negating such a unification, as required for disunification, yields universally quantified disjunctions that cannot be reduced to primitive disunification constraints. In these cases, the original optimization becomes unsound.

CLP(\set{}) systems such as \citet{dovierSetsAndCLP} address this by evaluating disjunctions directly through the host language's search. This approach is correct but can be inefficient when the disunification in question is purely structural. To recover the original performance in these cases, \proj{} adapts the technique described by \citet{FirstOrderMiniKanren}: rather than executing disjunctions immediately, \proj{} reifies them into the search tree as first-order terms. This representation enables the implementation to choose between different execution strategies based on the structure of the disjunction. In particular, the original two-watched literal optimization becomes a special case within a more general interpreter.

If all disjuncts are of the form \texttt{=/=} and all arguments are structural, \proj{} selects the watched-literal scheme: each disjunct is stored in the shared disjunction node, and only one variable per disjunct is watched. If this pattern can't be matched, the interpreter falls back to the general-purpose disjunction mechanism. This hybrid strategy ensures that performance is preserved for programs that rely only on structural disunification, while set disunifications remain sound and complete.

\subsection{Virtualizing \texttt{absento}}\label{subsec:virtualizing-absento}

The \texttt{absento} constraint in miniKanren has a subtle caveat when applied to lists: it recurs structurally on \texttt{cons} pairs as skewed binary trees rather than lists of elements. As a result, \texttt{absento} fails when its first argument matches any sublist of the second---even when that sublist is not an element of the list, but merely a suffix. For example, the following query fails:

\begin{alltt}
(run* (\_)
  (absento '(b c) '(a b c)))
\hook{} ()
\end{alltt}

This behavior is appropriate for \texttt{cons} pairs, which are often used to represent arbitrary trees with lists as a special case. However, this behavior is unsound in the case of sets, as it would expose internal details about how particular sets were constructed. If we allowed the same semantics we do for lists as we did for sets, the following queries would violate equational reasoning:

\begin{alltt}
(run* (\_)
  (absento '#(set (1)) '#(set (1) #(set (2)))))
\hook{} ((\_.0))
(run* (\_)
  (absento '#(set (1)) '#(set (2) #(set (1)))))
\hook{} ()
\end{alltt}

\noindent
In this case, two objects which would typically unify (as they represent the same set) cannot be used in place of one another. To solve this incongruity, we need to reconsider the semantics of \texttt{absento}. From a logical perspective, \verb|(absento p q)| posits two separate claims: \texttt{p} is not equivalent to \texttt{q}, and \texttt{p} is absent from all subterms of \texttt{q}. In traditional miniKanren, we can express the first constraint alone using \texttt{=/=} and both of them together using \texttt{absento}, but we can't express just the latter constraint. This isn't strictly necessary with the semantics that currently exist, but in the case we saw above, we require exactly that behavior.

To resolve this mismatch, \proj{} separates these two responsibilities. The constraint \texttt{absento} is removed entirely, and replaced with a new primitive \texttt{sub-absento} constraint that enforces the latter constraint alone: that \texttt{p} is absent from all subterms of \texttt{q}. We can see this in the following query:

\begin{alltt}
(run* (x)
  (absento '#(set) `#(set (1) ,x)))
\hook{} ((_.0 (set _.0) (sub-absento (#(set) _.0))))
\end{alltt}

It is worth noting that the cases in which this semantic definition of \texttt{absento} differs from the structural definition are very few, and although correct, separating \texttt{absento} into two constraints in most cases only serves to create extraneous constraints. To reduce this
overhead and keep compatibility with existing implementations, \proj{} re-introduces \texttt{absento} as a ``virtual constraint'': If both \texttt{(=/= p q)} and \texttt{(sub-absento p q)} are known to hold for \texttt{p} and \texttt{q} (either because they were directly asserted or implied by other constraints), the system displays these constraints as a virtual \texttt{absento} constraint during reification. This preserves compatibility with existing miniKanren idioms and tools, while respecting the new semantics of set objects. For example, consider the following query:

\begin{alltt}
(run* (p)
  (sub-absento 3 `#(set (1) ,p)))
\hook{} ((_.0 (absento (3 _.0))))
\end{alltt}

\noindent
In this case, we know that both the criteria for \texttt{absento} hold: \texttt{3} is absent from all members of the set by the explicitly listed constraint, and \texttt{3} cannot unify with \texttt{p}, as numbers and sets are disjoint types. As both the requirements hold, we're able to ``upgrade'' the constraint to \texttt{absento} without violating soundness.

\section{Limitations and Future Work}\label{sec:limitations}

The current implementation of \proj{} provides expressive constraints for reasoning about sets and association lists in miniKanren. However, several limitations remain, which suggest directions for future work.

\paragraph*{Runtime performance}\label{sec:limitations:runtime}

Set operations in \proj{} often trade soundness in worst-case scenarios for performance in best-case scenarios. Although sets are logically unordered and deduplicated, their underlying representation is list-based, which imposes linear-time costs for operations like membership and union. Future implementations may improve asymptotic performance through more efficient representations---such as balanced search trees---while preserving logical extensionality.

\paragraph*{Result duplication}\label{sec:limitations:duplicates}

While all set representations are semantically equivalent, their underlying structure can often be seen operationally. For example, consider the following query:

\begin{alltt}
(run* (p)
  (ino 2 `#(set (2 ,p 2))))
\hook{} (_.0 2 _.0)
\end{alltt}

\noindent
Despite the fact that the query is tautological, the set unification algorithm still searches each element individually. This redundancy in results, while always finite, can become arbitrarily large when several set relations are run in series. More disciplined constraint scheduling or structural rewrites could mitigate this leakage and give users finer control over operational behavior.

\paragraph*{Lack of modularity}\label{sec:limitations:modularity}

The constraints implemented by \proj{} are ``baked-in'' to the structure of the interpreter. Set and association list representations are fixed, and the implementation offers no mechanism to register or override constraint behavior. This makes it difficult to extend \proj{} to new domains or to experiment with application-specific data representations. Similarly, many constraints have logical implications for each other, but these interactions must be tracked by hand. For instance, if a key is known to be free in an association list, it must be distinct from all the keys the association list does bind:

\begin{alltt}
(run* (p q r)
  (freeo p r)
  (lookupo q r q))
\hook{} '(((_.0 _.1 _.2)
     (=/= ((_.0 _.1)))
     (lst _.2)
     (free (_.0 _.2))
     (lookup (_.1 _.2 _.1))))
\end{alltt}

\noindent
Not requiring the programmer to manually list each of these interactions---through constraints over tree structures, bounded universal quantifiers, or a declarative constraint-handling language---would reduce the burden of implementation and allow for further extensions written by users.

\paragraph*{Additional data-structures}
\proj{} constraints implement only a subset of a potentially more general data-structure framework. The system supports extensional, finite sets and association lists, but does not currently support multisets, lists, intensional sets, bounded universal quantification, dictionaries or other membership or finite mapping structures. Many of these features were explored by \citet{dovierUniformApproachToConstraintSolving, dovierIntegratingListsMultisetsAndSets} and could be incorporated into future versions of \proj{} with appropriate constraint definitions and solver support.

Addressing these limitations would improve both the expressiveness and predictability of \proj{}, and further support its use in interpreters, program synthesis, and other relational applications.

\section{Related Work}\label{sec:relatedwork}

\paragraph*{\{log\} \cite{dovierSetsAndCLP}} Dovier et al. originally introduced the CLP(\set{}) theory and authored the Prolog dialect \{log\} (read: set-log) which includes lazy set constraints. Their language supports reasoning about sets using a functionally complete constraint set, including membership, disjointness, and union. However, \{log\} is not relational: it incorporates extra-logical features such as \texttt{var/1} and relies on an incomplete depth-first search strategy. In contrast, \proj{} is built on top of miniKanren and inherits its complete interleaving search strategy and relational semantics.

\paragraph*{OCanren \cite{ocanren}} The OCanren system, a port of miniKanren to OCaml, supports user-defined Algebraic Data Types (ADTs), including those beyond traditional cons-based lists. However, these data structures are still subject to traditional miniKanren unification: that is, equality of data values is defined structurally. In contrast, \proj{} supports extensional equality over sets, as well as lazy constraints defined over them.

\paragraph*{core.logic \cite{corelogic}} The core.logic library for Clojure includes unification over maps, allowing reasoning about key-value associations. However, its unification strategy is limited: keys must be ground symbols, and many desirable constraints---like operations on partially instantiated dictionaries or lazy dictionary operations---are not supported. In contrast, \proj{} provides lazy association list constraints that support arbitrary keys.

\section{Conclusion}\label{sec:conclusion}

We have presented an extension to miniKanren that adapts the first-class relational treatment of sets from CLP(\set{}) and generalizes it to support association lists. These data structures improve the expressiveness of relational programs and offer better failure behavior in interpreters and other applications where partial information is common. Our implementation demonstrates how such structures can be built atop a modern miniKanren system without requiring changes to its core.

We hope this work encourages further exploration into the design and implementation of data structures optimized for relational programming. By making structures like sets and maps first-class, relational programmers gain tools better suited to reasoning about incomplete, symbolic, or evolving data.

We would like to thank Michael Ballantyne for both his work on faster-miniKanren and his help on the design of \proj{}. We would also like to thank Anastasia Kravchuk-Kirilyuk for her feedback on this paper.

\bibliographystyle{ACM-Reference-Format}
\bibliography{paper}

\end{document}